\documentclass[sigconf]{acmart}

\theoremstyle{definition} 
\newtheorem{finding}{Finding}

\usepackage{multirow}
\usepackage{listings}
\usepackage{xcolor}
\usepackage{subcaption}
\usepackage{enumitem}

\AtBeginDocument{%
  }

\copyrightyear{2026}
\acmYear{2026}
\setcopyright{cc}
\setcctype{by}
\acmConference[LLM4Code '26]{3rd International Workshop on Large Language Models For Code }{April 12--18, 2026}{Rio de Janeiro, Brazil}
\acmBooktitle{3rd International Workshop on Large Language Models For Code (LLM4Code '26), April 12--18, 2026, Rio de Janeiro, Brazil}
\acmPrice{}
\acmDOI{10.1145/3786181.3788710}
\acmISBN{979-8-4007-2412-1/2026/04}

\begin{document}    

\title{RubberDuckBench: A Benchmark for AI Coding Assistants}

\author{Ferida Mohammed}
\affiliation{%
  \institution{Bryn Mawr College}
  \city{Bryn Mawr}
  \country{USA}}
\email{fmohammed@brynmawr.edu}

\author{Fatma Ayad}
\affiliation{%
  \institution{Bryn Mawr College}
  \city{Bryn Mawr}
  \country{USA}}
\email{fayad@brynmawr.edu}

\author{Petros Maniatis}
\affiliation{%
  \institution{Google DeepMind}
  \city{Mountain View}
  \country{USA}}
\email{maniatis@google.com}

\author{Satish Chandra}
\affiliation{%
  \institution{Meta Platforms}
  \city{Menlo Park}
  \country{USA}}
\email{schandra@acm.org}

\author{Elizabeth Dinella}
\affiliation{%
  \institution{Bryn Mawr College}
  \city{Bryn Mawr}
  \country{USA}}
\email{edinella@brynmawr.edu}


\begin{abstract}
Programmers are turning to AI coding assistants to answer questions about their code. Benchmarks are needed to soundly evaluate these systems and understand their performance. To enable such a study, we curate a benchmark of real-world contextualized questions derived from Github pull request comments. Out of this work, we present RubberDuckBench: a multilingual benchmark of questions about code, along with detailed rubrics for evaluating answers. We evaluate a diverse set of 20 LLMs (proprietary \& open-source) on answering these questions. We find that even state of the art models fail to give consistent, correct responses across the benchmark. Grok 4 (69.29\%), Claude Opus 4 (68.5\%), and GPT-5 (67.8\%) perform best overall, but do not exhibit pairwise significant superiority over the next 9 best performing models. Most models obtain points through partial credit, with the best performing models only answering at most 2 questions completely correctly across all trials. Furthermore, models often hallucinate with lies in 58.3\% of responses on average. Cost analysis reveals no correlation between expense (API pricing or parameter count) and performance. We intend this benchmark to be a target for future research in trustworthy and correct AI coding assistants. 
\end{abstract}



\maketitle

\section{Introduction}
Software engineering has been revolutionized by AI systems at nearly every stage of the developer workflow~\cite{llm-se-survey}. Popular AI enabled IDEs~\cite{copilot, tabnine, alphacode, cursor} are typically configured with a chosen LLM and allow for AI driven code generation, repair, and search. These systems are also equipped with a chat functionality. Studies have shown that AI tools increase productivity and perceived productivity in development tasks which require code writing, editing, and search~\cite{google-speed,ziegler}. However, the in-context chat functionalities of these systems are understudied. Programmers are increasingly turning to AI coding assistants to answer questions about their code, with ``searching for answers'' listed as the most common AI use case according to the Stack Overflow 2025 Developer Survey~\cite{stack-overflow-survey}. Benchmarks to evaluate this increasingly popular component of the AI driven developer workflow are needed. 

Predominant benchmarks for evaluating LLMs trained on code primarily target code generation from natural language descriptions~\cite{codex, MBPP}, and involve generating standalone functions. Other benchmarks target code generation at a larger scale in more specific developer settings such as secure backend development~\cite{baxbench}, or web development~\cite{webbench}. Beyond code generation, benchmarks have been proposed for various tasks in the development workflow such as translation~\cite{codexglue}, repair~\cite{swebench,repairBench}, and auditing~\cite{scbench}. None of these benchmarks evaluate the use case of asking contextualized coding questions. The most similar benchmarks to this setting~\cite{stackeval, robustAPI} are derived from Stack Overflow questions. Although these indeed provide a target of real world coding questions, they offer a different out-of-context perspective. Stack Overflow questions are typically general language level questions that don't require deep reasoning over project specific code. No existing benchmarks directly evaluate an LLMs ability to answer questions about contextualized code. 

To derive such a benchmark, we turn to Github pull request comments. Unlike Stack Overflow questions, these comments are contextualized within a given project, file, and line. They provide a realistic parallel to the scenario of a developer asking contextualized questions to an AI coding assistant. However, they cannot simply be mined as targets for this benchmark. Pull request comments are often phrased as code review style edit suggestions rather than queries and are not always suitable to ask an AI coding assistant. Through a study of the CodeReview~\cite{codereviewer} dataset of pull request data from high-quality open-source repositories, we find that comments discuss a wide variety of topics. At various densities between languages, conversations span specification refinement, to shallow edit suggestions, to discussions between author and reviewer reasoning over the given code. The questions in our benchmark are derived from comments in the latter category. They are classified as such and rephrased as questions through an LLM-human annotation approach. Viable responses to these questions are not unique, and can be correctly phrased in many ways. They are open-ended and nuanced in the amount of information that is relevant. To properly evaluate a given response to these questions, we manually curate detailed rubrics.  

Through this, we present RubberDuckBench: a benchmark for AI coding assistants. Our benchmark contains 15 contextualized questions, with rubrics for evaluating answers. The questions were derived from 13 open source Github projects, with an average of 25.3k stars, and are evenly split between Java, Python, and C++. Questions in our benchmark are contextualized to specific locations within files and projects. They address: the behavior of in-project code, how library code functions and its implications in the context, performance considerations, and the values of program variables. Each question is concise, unambiguous, and grounded in the code context. 


With this benchmark, we study whether state-of-the-art LLMs can correctly answer questions about contextualized code.  We evaluate 20 LLMs on RubberDuckBench, studying their overall ability, cost, and propensity to lie. We find that, in general, models struggle to provide consistent correct responses across the benchmark. Grok 4~\cite{grok-4} (69.29\%), Claude Opus 4~\cite{claude-4} (68.53\%), and GPT-5~\cite{gpt-5} (67.80\%) perform best overall, but do not exhibit pairwise significant superiority over the next 9 best performing models. Most models obtain points through
partial credit, with the best performing models only answering at most 2 questions completely correctly across all trials. Furthermore, models often hallucinate with lies in 58.3\% of responses on average. Cost analysis reveals no correlation between expense (API pricing or parameter count) and performance. 

In summary, we contribute:
\begin{enumerate}
    \item A multilingual benchmark of real-world, contextualized questions that require code reasoning to answer. 
    \item An evaluation of 20 models on our benchmark and discussion of findings. 
    \item A reproducible evaluation package\footnote{https://cs.brynmawr.edu/RubberDuckBench}.
\end{enumerate}
We intend this benchmark to act as a target for further research into trustworthy and correct AI coding assistants. 



 \section{Motivation}


Software engineering has been transformed by AI systems at various stages in the software development workflow~\cite{llm-se-survey}. AI enabled developer tooling platforms such as Cursor~\cite{cursor}, Github Copilot~\cite{copilot}, and others~\cite{tabnine, kiro}, offer various neural functionalities including code completion, search, and chat, parameterized by an LLM of the user's choosing. Studies have shown that AI tools increase productivity and perceived productivity in development tasks which require code writing, editing, and search~\cite{google-speed, ziegler}. However, the chat functionality of these tools is understudied and increasingly leveraged. According to the Stack Overflow 2025 developer survey, the most common use of AI is to seek answers to questions about their code~\cite{stack-overflow-survey}. Despite its popularity, it is unclear how well models perform at answering contextualized questions. Do state of the art LLMs answer questions about contextualized code correctly? Do they hallucinate or lie about API or project specific facts? Do certain models perform better than others? In this paper, we aim to answer these questions. 

The predominant benchmarks for evaluating LLMs trained on code are HumanEval~\cite{codex} and MBPP~\cite{MBPP}. They measure a model’s ability to synthesize programs from natural language docstrings. These text-to-code benchmarks are synthetic, handwritten, and involve generating a standalone function. Other, more realistic benchmarks, target code generation at a larger scale in more specific developer settings. BaxBench~\cite{baxbench} evaluates a model's ability to generate correct and secure backend applications. Web-Bench~\cite{webbench} measures a model's ability to perform realistic, sequential, full stack development tasks. Beyond code generation, benchmarks have been proposed for various tasks in the developer workflow. SWE-Bench~\cite{swebench}, a widely used benchmark, measures a model's ability to resolve a given issue by modifying the codebase. CodeXGLUE~\cite{codexglue} is a benchmark for general code understanding and generation. Beyond generation, it includes sub-benchmarks for translation between programming languages, code search, clone detection, defect detection, and code summarization. Benchmarks have also been proposed for post-development scenarios such as auditing. SC-Bench~\cite{scbench} measures a model's ability to detect violations of standards on the Ethereum blockchain. Other works~\cite{cve}, measure a model's ability to statically detect CVE and CWEs in a given project. These benchmarks fundamentally target a different task than the task we aim to study: answering non-trivial questions about contextualized code.  


The most similar works to ours are derived Stack Overflow questions. StackEval~\cite{stackeval} and RobustAPI~\cite{robustAPI} include benchmarks of real world questions asked on Stack Overflow. Although these indeed provide a target of real world coding questions, they offer a different out-of-context perspective. Stack Overflow questions are typically
general language level questions that don’t require deep reasoning over project specific code. For example, consider the following question from StackEval: \textit{"In Java, when should we use private instance methods in interfaces?"}. In contrast, the questions we aim to study are ones which are asked in an AI enabled IDE chat window. They are contextualized in a project, and typically require some level of reasoning over the given code to answer. 
\section{Benchmark Curation}
\label{sec:curation}
In this work, we aim to study an LLMs ability to answer real world, contextualized, questions about code. To derive such a benchmark, we turn to Github pull request comments. Unlike Stack Overflow questions, these comments are contextualized within a given project, file, and line. They provide a realistic parallel to the scenario of a developer asking contextualized questions to an AI coding assistant. To mine pull request comments, we leverage the CodeReview~\cite{codereviewer} dataset
contains pull request data from publicly available high-quality open-source repositories. 

\begin{figure}[h]
\centering
\setlength{\abovecaptionskip}{2pt}
\setlength{\belowcaptionskip}{2pt}
\includegraphics[width=0.48\textwidth]{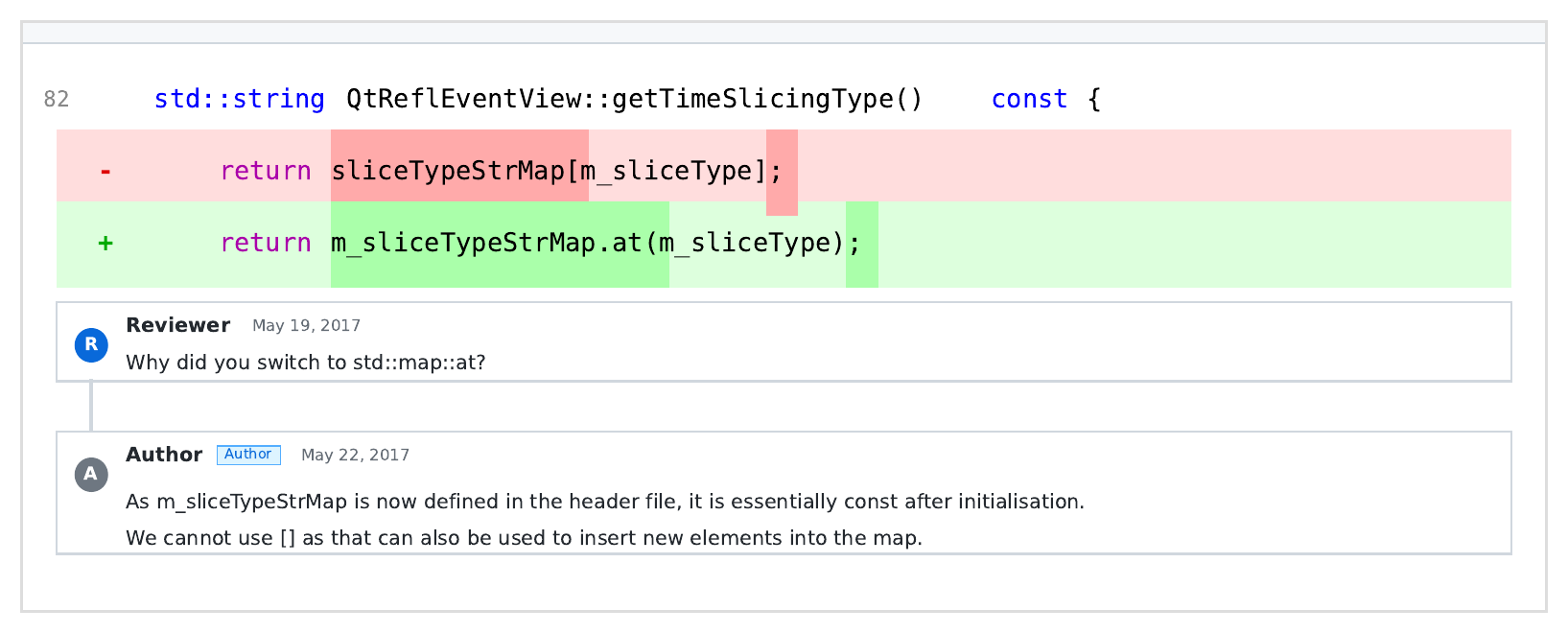}
\caption{\footnotesize PR comment exemplifying a contextualized question.}
\label{fig:code-review-comment}

\vspace{1em}  

\begin{minipage}{0.45\textwidth}
\begin{lstlisting}[language=C++, basicstyle=\ttfamily\footnotesize, keywordstyle=\color{blue}, commentstyle=\color{green!60!black}, stringstyle=\color{red}, frame=single, breaklines=true]
string QtEventView::getTimeSlicingType() const {
  return m_sliceTypeStrMap.at(m_sliceType);
  /*Question: Is there a difference between using 
                m_sliceTypeStrMap.at(m_sliceType) vs 
                m_sliceTypeStrMap[m_sliceType]?*/
}
\end{lstlisting}
\end{minipage}
\caption{\footnotesize Rephrased PR comment as a question for an AI Coding Assistant.}
\label{fig:code}
\Description{In the original PR comment, the code reviewer asks: "Why did you switch to std::map::at?". The author replies: "As m_sliceTypeStrMap is now defined in the header file, it is essentially const after initialization. We cannot use [] as that can be used to insert new elements into the map." This comment is rephrased as "Is there a difference between using m_sliceTypeStrMap.at(m_sliceType) vs m_sliceTypeStrMap[m_sliceType]?"}
\end{figure}

We first study the CodeReview dataset to understand the form and subject of pull request comments. In terms of subject, we find that pull request comments do provide a target of real world contextualized queries, but many comments are unsuitable to ask an AI coding assistant (e.g. discussions about functional behavior or planning between members of the development team). In terms of form, we find that pull request comments are rarely phrased as concise yet unambiguous questions. Consider Figure~\ref{fig:code-review-comment} where an anonymized reviewer asks a question about the use of \texttt{std::map::at} vs the use of \texttt{operator[]} in the context of the changed code. The anonymized PR author responds that the \texttt{std:map:at} call is used due to the \texttt{const} context. The subject of this comment is suitable to ask an AI coding assistant, but certainly not in its current form. A rephrased version of this query can be found in Figure~\ref{fig:code}. The question captures the underlying query from the reviewer, but rephrases it to avoid referencing the edit, and grounds the question in concrete program elements. In this section, we describe our LLM-human annotation approach to filter and rephrase pull request comments to a target set of concise and unambiguous questions. Lastly, we describe our approach for collecting answers to these questions. Viable responses are not unique and can be correctly phrased in many ways. They are open-ended and nuanced in the amount of information that is relevant. To properly evaluate a given response to our questions, we manually curate detailed rubrics. In the following subsections we present our study of pull request comments in the CodeReview dataset, propose an LLM assisted approach for question derivation, and describe our manual technique for rubric curation.


 
\vspace{-2em}

\subsection{Study of Pull Request Comments}
In this section, we present a study of pull request comments in the CodeReview dataset. We limit our analysis to our three target languages: Java, Python, and C++. For each language, we manually analyze 100 randomly selected pull request comments and categorize them into three types based on their subject matter: code reasoning comments, specification discussions, and shallow edit suggestions. Shallow edit suggestions involve refactoring recommendations, planning or documentation comments, or edit suggestions related to naming conventions or code formatting. Specification discussions involve threads between collaborators refining the intended program behavior. Lastly, code reasoning comments involve discussions that require reasoning over control-flow or value propagation. We derive the questions in our benchmark from this latter category.

\begin{table}[h]
\centering
\begin{tabular}{l|ccc}
\hline
 & Java & Python & C++ \\
\hline
Code Reasoning Comment & 23\% & 49\% & 35\%\\
Specification Discussion & 10\% & 11\% & 30\% \\
Shallow Edit Suggestion & 67\% & 40\% & 35\% \\
\hline
\end{tabular}
\caption{Pull Request Comments By Type.}
\label{tab:review_comments}
\end{table}

Table~\ref{tab:review_comments} presents the breakdown of the 100 pull request comments for each language. Code reasoning comments appear at varying densities across the three languages. However, for all languages, the majority of pull request comments consist of specification discussions and shallow edit suggestions. Through this study, we find that pull request comments provide a target of real-world contextualized queries about code, but they are not always phrased as questions, and often discuss topics that are not suitable to ask an AI coding assistant. 

\subsection{LLM Assisted Question Curation}
Motivated by the high density of pull request comments that discuss topics which are not suitable for our benchmark, we develop an LLM-assisted approach to filter the CodeReview dataset. First, we query an LLM to rephrase a given code review comment as a concise question that someone familiar with the codebase could understand and potentially answer. Then, we prompt the LLM to determine if the rephrased question is suitable to ask an AI coding assistant. All LLM outputs are verified and potentially further rephrased by human annotators. Our LLM-assisted annotation technique is parameterized with Claude Opus 4.1 as we empirically found it to have better performance than other flagship models, and produced more stable outputs across multiple trials. We also experimented with performing filtering before rephrasing, but found that LLMs performed more accurate classification over rephrased questions rather than the original code review comment.

Given the overall large quantity of pull request comments and our intention to leverage the LLM system as an annotation assistant rather than retrieve all relevant comments, we evaluate our system on precision. Our approach  achieved an average precision of .78 with slightly higher performance on Python (.84) than Java (.79) and C++ (.71). We also experimented with a keyword matching approach, but found it to be too rigid and unable to recognize nuanced discussions with a precision of .39. We apply the LLM-assisted technique, parameterized with Claude, on the CodeReview dataset to retrieve questions flagged as suitable to ask an AI Coding Assistant. We manually review, requiring agreement across 3 authors, and further rephrase when necessary. For each language, we select the first 5 suitable questions from this output for inclusion in our benchmark. 
\vspace{-1em}


\begin{figure*}
\small
\centering
\begin{tabular}{|p{0.15\linewidth}|p{0.7\linewidth}|p{0.05\linewidth}|}
\hline
\textbf{Criterion} & \textbf{Description} & \textbf{Points} \\
\hline
\hline
\multirow{2}{0.2\textwidth}{Difference \\ between operations} 
& Does the answer state or imply that there is a difference between using \texttt{at} vs \texttt{[]}? & 2 \\
\cline{2-3}
& \textbullet~Deduct 1 point if the answer does not state either way. & \\
& \textbullet~Deduct 2 points if the answer incorrectly states there is no difference. & \\
\hline
\multirow{2}{0.2\textwidth}{Non-existent\\ key handling} 
& Does the answer discuss how each operation handles non-existent keys? & 3 \\
\cline{2-3}
& \textbullet~If the answer does not mention that there is a difference in each operation’s handling of non-existent keys, 1 point should be deducted.  & \\
& \textbullet~The answer should mention that for this code, the map will never be queried with a non-existent key. If this is not mentioned, 1 point should be deducted. & \\
& \textbullet~If the answer states or implies that non-existent keys are possible by listing it as an important consideration in this context, 2 points should be deducted.  & \\
\hline
\multirow{2}{0.2\textwidth}{Usage in const methods} 
& Does the answer discuss how each operation can be used in the context of a \texttt{const} method? & 2 \\
\cline{2-3}
& \textbullet~The answer should mention or imply that the method is marked as \texttt{const}. Deduct 1 point if not mentioned. & \\
& \textbullet~The answer should explain that \texttt{[]} cannot be called in a \texttt{const} method. Deduct 1 point if not mentioned. & \\
\hline
\hline
\multicolumn{2}{|r|}{\textbf{Total Points:}} & \textbf{7} \\
\hline
\end{tabular}
\caption{Rubric: Difference Between Contextualized \texttt{at} vs \texttt{[]} Operators}
\label{fig:rubric}
\Description{The rubric contains criterion for differences between the operators, handling of non-existent keys, and usage in const methods.}
\end{figure*}
\subsection{Rubric Curation}
\label{sec:curation:rubric}
Viable answers to RubberDuckBench questions are not unique and can be correctly phrased in many ways. For example, the question in Figure~\ref{fig:code} could be answered equivalently with \textit{Yes, there is a difference if m\_sliceType is not in the structure... } or \textit{No, there is no difference if m\_sliceType is in the structure...}. As such, syntactic metrics that compare responses to a gold standard answer are insufficient. To ensure an accurate evaluation that captures nuanced definitions of correctness, we manually curate rubrics which can be applied to diverse open-ended LLM responses. 

To develop our rubrics, we first carefully study each project, pull request, and comment discussion thread. We create minimal executable examples of each relevant code segment and write test cases demonstrating the correct answers. We include these ``proof'' scripts in our evaluation package. Once we are confident in our understanding of the given question and codebase, we curate a rubric that can be applied fairly at scale through the following process. Three authors, including an Assistant Professor, were involved in rubric creation and trained in fair rubric development. Each rubric curator participated in both individual drafting and group discussion. After reaching consensus, we iteratively refine rubrics using methods inspired by the College Board for Advanced Placement exams which evaluate diverse student responses at scale~\cite{apcs}. Based on answers from a ``training set'' of randomly selected model responses, we iteratively update the rubric to ensure comprehensive coverage of possible answers. Each rubric took an average of 12 person hours to curate and refine. 

Rubrics contain high-level criteria covering essential information that correct responses should include. Rubric sub-criteria penalize hallucinations (incorrect or fabricated information) more heavily than omissions of relevant information. All rubrics follow a negative-scoring scheme in which responses begin with a perfect score and points are deducted for errors. Figure~\ref{fig:rubric} shows the rubric for our motivating example (PR comment in Figure~\ref{fig:code-review-comment} and rephrased question in Figure~\ref{fig:code}).




\section{RubberDuckBench}
\label{sec:benchmark}
Our benchmark contains 15 questions evenly split between each language: Java, Python, and C++. Each question is contextualized in a particular project, git commit, and line number. Our artifact includes scripts to automatically clone and checkout the necessary context. Each question also includes a detailed rubric curated through the process discussed in Section~\ref{sec:curation:rubric} and minimal script which exemplifies the answer. All samples are given in our artifact. Due to the significant manual effort required for rubric curation and and cost to run flagship proprietary models, RubberDuckBench is small and functions as a target for a deep case study. 
\vspace{-.2em}
\begin{table}[h!]
\centering
\setlength{\abovecaptionskip}{3pt}
\setlength{\belowcaptionskip}{3pt}
\small
    \begin{tabular}{l|l|l}
    \hline
        \textbf{Question} & \textbf{Question Type} & \textbf{Project} \\ 
            \hline
        Java 1 & Project behavior & Mozilla  / thunderbird-android \\ 
        Java 2 & Library behavior & Mozilla  / thunderbird-android \\ 
        Java 3 & Value & Pinterest / Secor \\ 
        Java 4 & Value & Alibaba / Nacos \\ 
        Java 5 & Project behavior & AntennaPod / AntennaPod \\ 
        Python 1 & Library behavior & DMLC / DGL \\ 
        Python 2 & Project behavior & BlazeMeter / Taurus \\ 
        Python 3 & Library behavior & PSF / Requests \\ 
        Python 4 & Library behavior & Hyperledger / indy-node \\ 
        Python 5 & Value & Google / ClusterFuzz \\ 
        C++ 1 & Performance & Cuberite / Cuberite \\ 
        C++ 2 & Value & PyTorch / PyTorch \\ 
        C++ 3 & Library behavior & Mantid Project / Mantid \\ 
        C++ 4 & Performance & Microsoft / Terminal \\ 
        C++ 5 & Project behavior & Mantid Project / Mantid \\ 
    \hline
    \end{tabular}
    \caption{RubberDuckBench Question Types and Projects.}
    \label{tab:benchmark}
\end{table}

In Table~\ref{tab:benchmark}, we present a taxonomy of our benchmark by question type. Project Behavior questions are about the functionality of in-project code. Value questions are about the value of program variables and how they propagate through the code context. Performance questions are about efficiency and performance considerations. Library Behavior questions are about the functionality of library or API code, yet are different from Stack Overflow style questions as they ask about behavior in the context of the given project. Our motivating example (Figures~\ref{fig:code-review-comment} and~\ref{fig:code}) is a Library Behavior question as it asks about the \texttt{std::map} API, with implications in the \texttt{const} context. We also list the popular open source project each data point was drawn from, averaging 25.3k stars per project.


\section{Evaluation}
In this section, we present our evaluation of state of the art LLMs on  RubberDuckBench. We answer the following research questions:
\begin{description}[leftmargin=20pt, labelindent=20pt]
\item[RQ1:] How do LLMs perform on RubberDuckBench?
    \item[RQ2:] What are the resource-performance tradeoffs?
    \item[RQ3:]  How frequently do LLMs hallucinate?
\end{description}

\noindent\textbf{Model Selection.} For a comprehensive evaluation, we select models from 8 popular providers. For each provider, we target models released in 2025. If a provider releases multiple classes of models (e.g. reasoning, budget, open source, code specific), we target models from each class. For open source models, we target the largest model we can run on our available compute. This selection criteria results in a target evaluation set of 20 LLMs. In this work, we evaluate Anthropic's state of the art reasoning models (Claude Opus 4.1~\cite{claude-opus-4.1} and Claude Opus 4~\cite{claude-4}) as well as their lower cost non-reasoning models (Claude Sonnet 4~\cite{claude-4} and Claude Sonnet 3.7~\cite{claude-sonnet}). We also evaluate a diverse set of OpenAI models including reasoning models GPT-5~\cite{gpt-5} and o3~\cite{gpt-o3}, non-reasoning model GPT-4.1~\cite{gpt-4.1}, and their open source models Gpt-oss-120B~\cite{gpt-mini-120}, and Gpt-oss-20B~\cite{gpt-mini-20}. In Google's family of models we evaluate Gemini 2.5 Pro~\cite{gemini-pro}, and their lower cost, higher speed models (Gemini 2.5 and 2.0 Flash~\cite{gemini-2.5-flash}). For the providers xAI, Alibaba, and Meta we evaluate one reasoning model and one non reasoning model (respectively) from each provider: Grok 4~\cite{grok-4} and Grok 3~\cite{grok-3}, Qwen3 and Qwen3 Coder~\cite{qwen3}, Llama 3.3 70B~\cite{llama3} and Llama 4 Scout~\cite{llama4}. Lastly, we include reasoning model Deepseek-R1 70B~\cite{deepseek} and non-reasoning model Mistral Large~\cite{mistral} in our evaluation. This comprises a diverse target evaluation set of 11 reasoning models, 4 models advertised as low cost, 12 proprietary models, and 8 open source models with sizes ranging from 8B to 123B. 
\\

\noindent\textbf{Experimental setup.} 
To elicit high quality responses, each model is prompted using best practices from recent prompt engineering research~\cite{prompt-engineering}. Notably, a system prompt is given to adopt a clear developer persona, deeper reasoning is enabled through Chain-of-Thought prompting~\cite{CoT}, and input and output formats are standardized with HTML-style tags. Open source models were run locally on an NVIDIA H100 GPU and proprietary models were run through their API endpoints. All models were run with a temperature of 0.01 for near deterministic results. Three authors independently reviewed all model responses and manually applied the evaluation rubrics, achieving high inter-rater reliability (ICC3 = .991) \cite{shrout1979intraclass}. Although related works use LLMs as a judge for evaluation, we found this to not be as reliable, with an ICC3 of 0.709 amounting to an error of over 20\% in some cases. Each model was evaluated three times per question, to account for potential non-determinism, and we report the average score. 

\subsection{RQ1: Performance on RubberDuckBench}
\begin{figure*}[t]
\centering
\begin{subfigure}[b]{0.47\textwidth}
    \includegraphics[width=\textwidth]{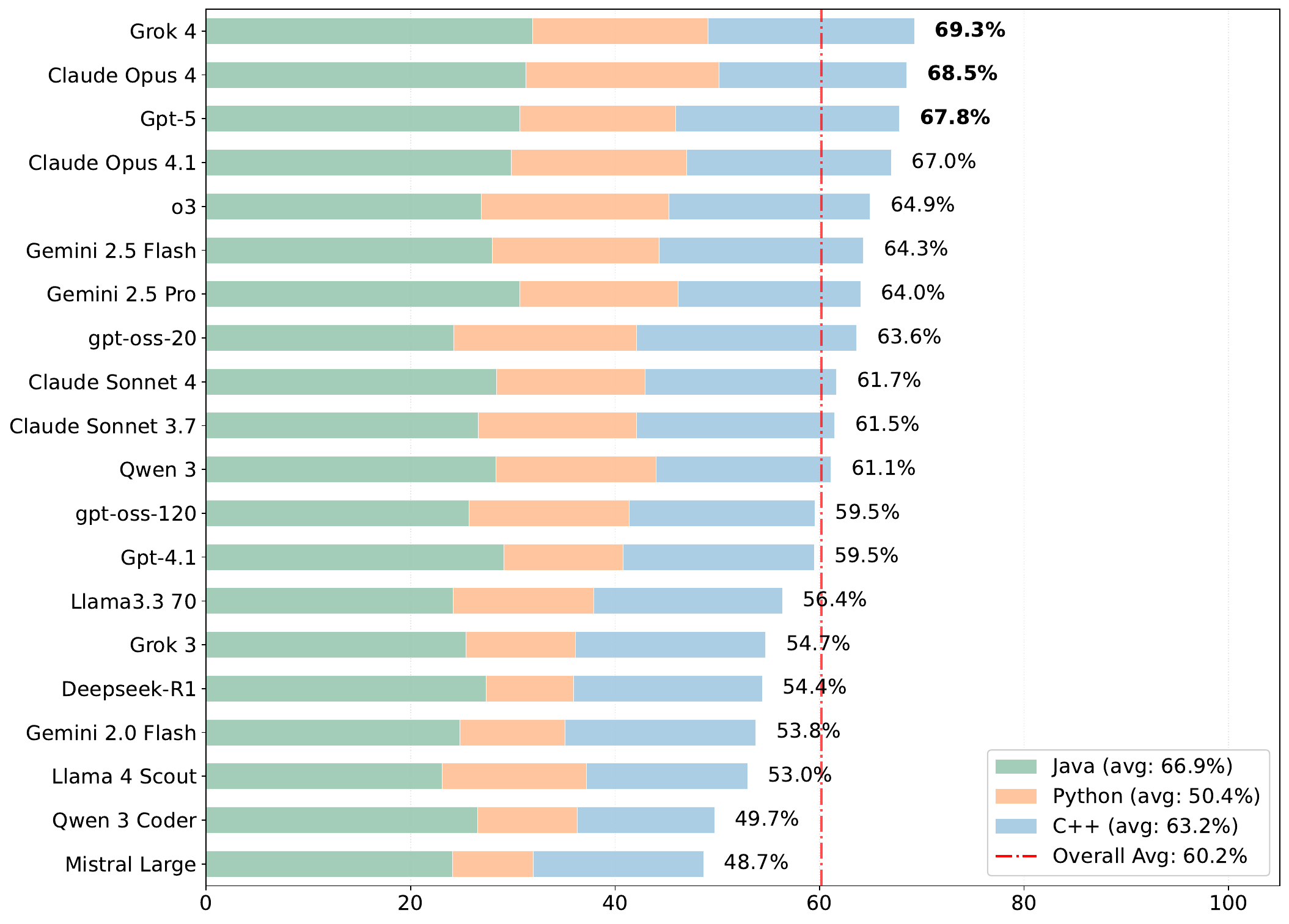}
    \caption{Average RubberDuckBench Score Across 3 Trials.}
    \label{fig:overall}
\end{subfigure}
\hfill
\begin{subfigure}[b]{0.47\textwidth}
    \includegraphics[width=\textwidth]{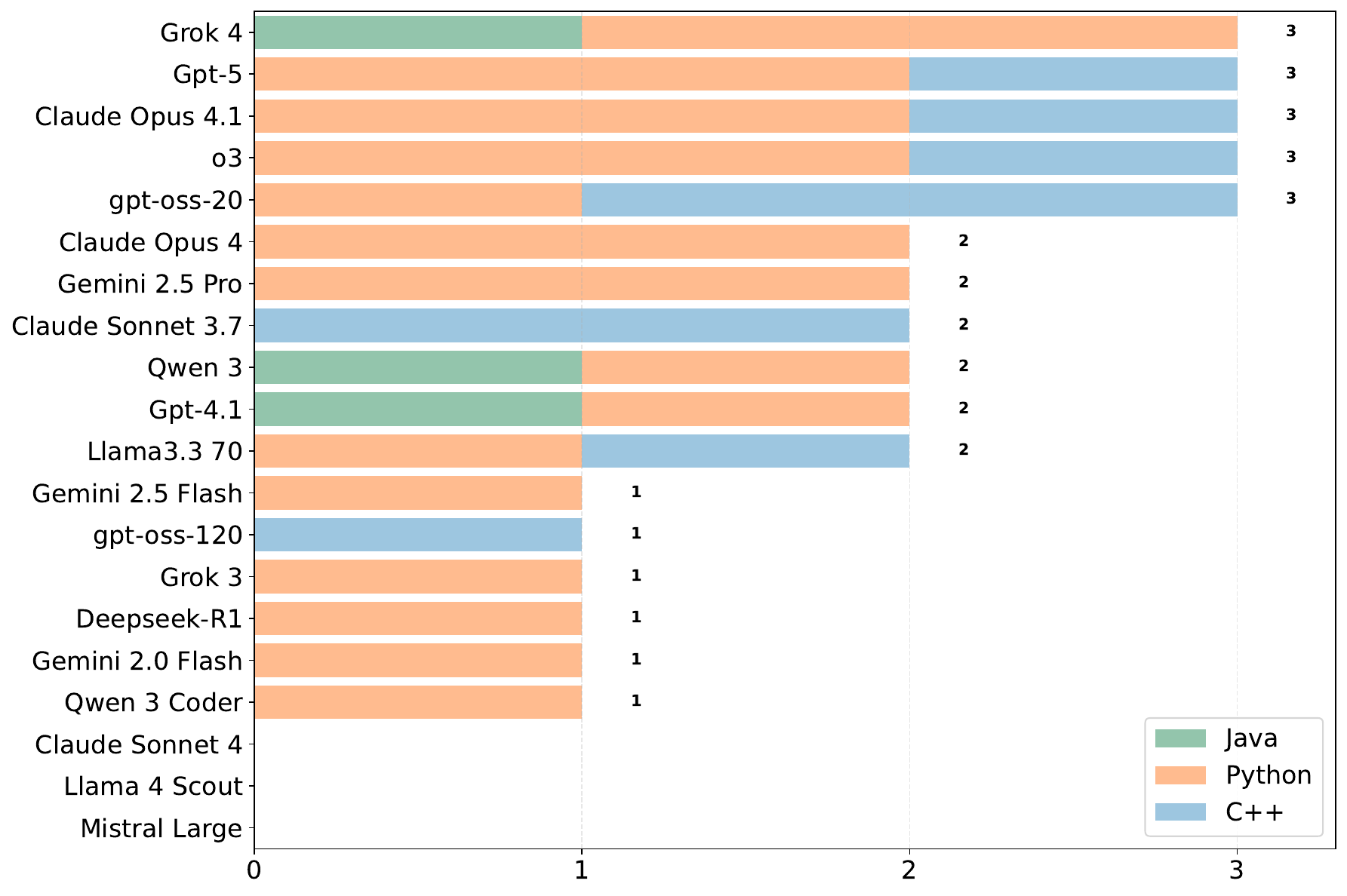}
    \caption{Number of Questions Answered Completely Correctly.}
    \label{fig:binary}
\end{subfigure}
\begin{subfigure}[t]{0.49\textwidth}
    \includegraphics[width=\textwidth]{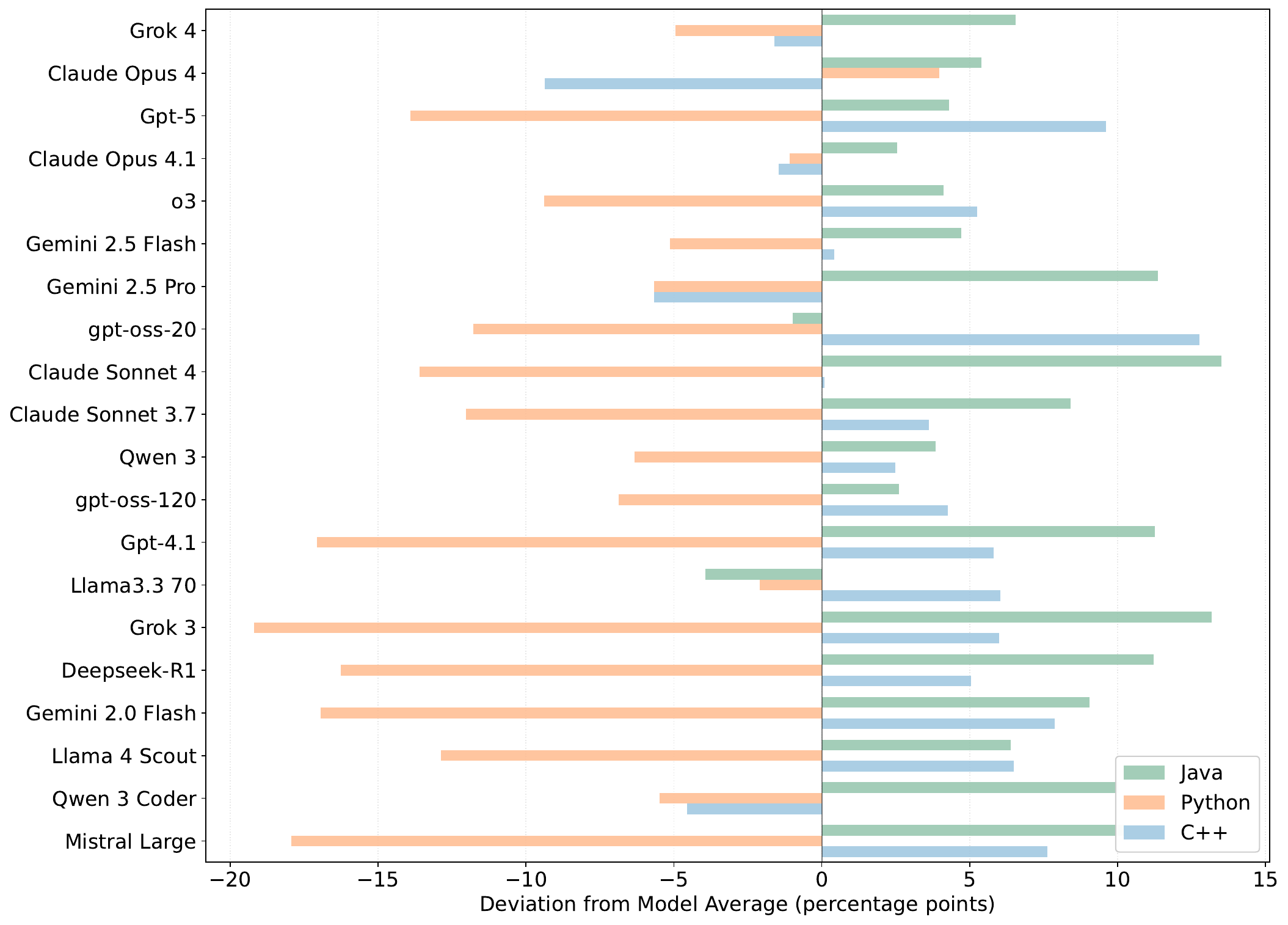}
    \caption{Deviation of Performance Across Languages.}
    \label{fig:lang}
\end{subfigure}
\hfill
\begin{subfigure}[t]{0.44\textwidth}
    \includegraphics[width=\textwidth]{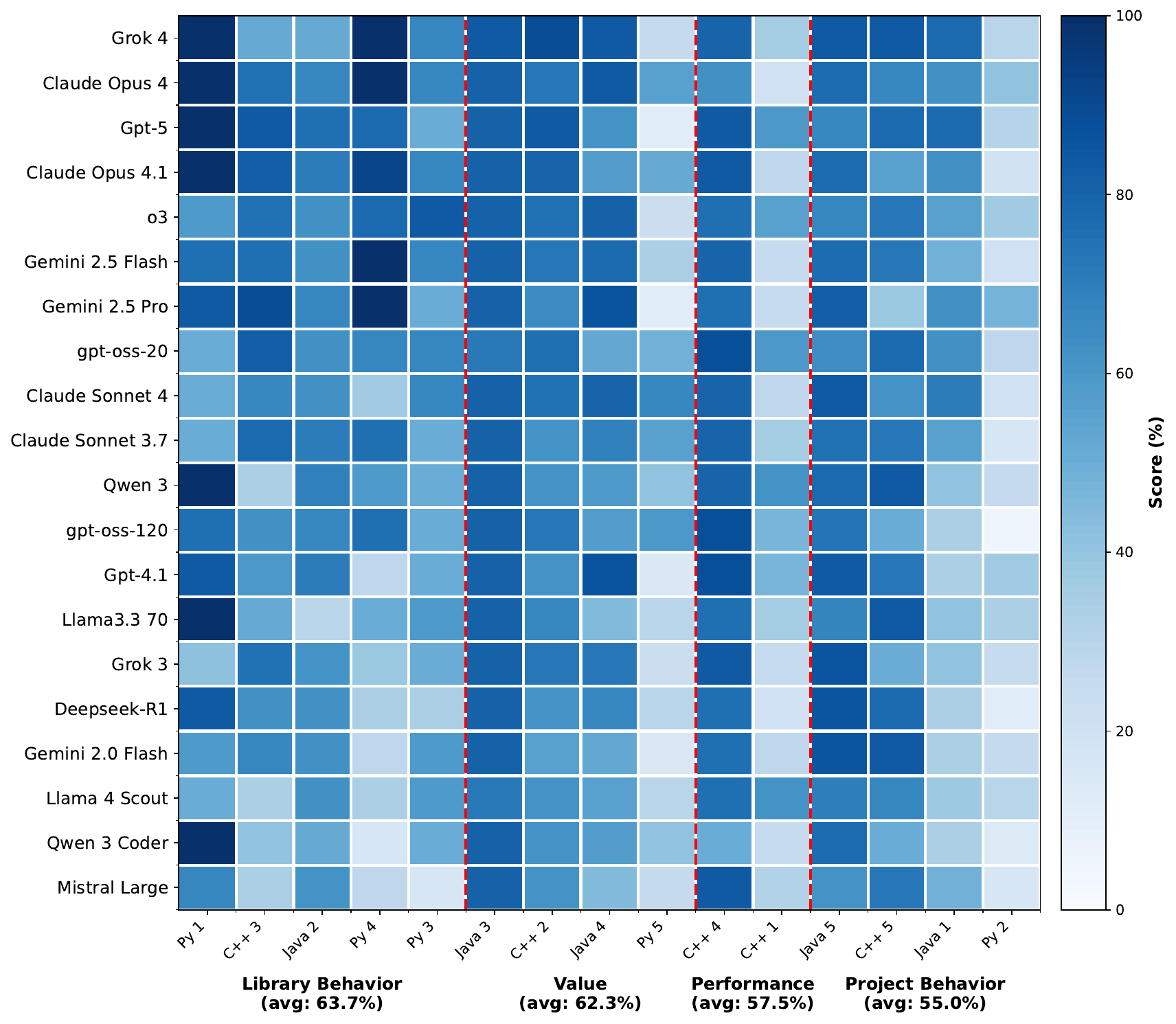}
    \caption{Performance Across Question Types.}
    \label{fig:questiontype}
\end{subfigure}
\caption{RQ1: Analysis of Model Performance}
\label{fig:combined}
\Description{Figure 4a shows a bar graph of the average performance of each model. Grok-4 performs the highest with 69.3\%. Figure 4b shows a bar graph of the total number of questions answered completely correctly. Grok-4, Gpt-5, Claude Opus 4.1, o3, and gpt-oss-20 perform best with 3. There is a breakdown by language. Figure 4c shows a bar graph breakdown of performance across programming languages. Figure 4d shows a heatmap of performance per question type. Models perform best on Library Behavior questions and worst on Project Behavior questions.}
\end{figure*}

In this section we answer the question: How do LLMs perform on answering questions in RubberDuckBench? We report our main findings in Figure~\ref{fig:overall}. The average and median scores were 60.17\% and 61.30\%. Grok 4 performs best overall with a score of 69.29\%. The next best performing models are Claude Opus 4 (68.53\%), GPT-5 (67.80\%), Claude Opus 4.1 (67.02\%), o3 (64.93\%), and Gemini 2.5 Flash (64.30\%). Notably, Claude Opus 4 performed better than its successor Claude Opus 4.1, and Gemini 2.5 Flash, a model that Google advertises as low-cost, performed better than its higher cost alternative Gemini 2.5 Pro. We also evaluated open source models with the best performing to be Gpt-oss-20 (63.63\%), performing better than Gpt-oss-120 (59.54\%), with 100 billion less parameters. The worst performing models were Mistral large (48.67\%), Qwen 3 Coder (49.73\%), Llama 4 Scout (52.96\%), and Gemini 2.0 Flash (53.78\%). 

In general, model performance varied substantially across individual questions. Although Grok 4 achieved the highest overall performance, lower-ranked models frequently outperformed it on particular questions. Figure~\ref{fig:questiontype}  shows a heatmap of each model's performance across the 15 questions. To determine whether higher aggregate performance translates to consistent pairwise superiority, we computed p-scores between models. Grok 4 only significantly outperformed (p < .05) the 7 lowest-ranking models. For the remaining 12 models, including Claude Opus 4, GPT-5, Gemini 2.5 Flash, and Gpt-oss-20, p-scores exceeded .05 when compared to Grok 4. This indicates that strong overall performance does not indicate superiority on individual benchmark questions, even against models with lower aggregate scores.

\begin{finding}
Models averaged 60.17\% (median: 61.30\%) across RubberDuckBench. Grok 4 achieved the highest overall score (69.29\%), but did not consistently outperform other top models, showing no significant differences (p < .05) from the next 12 best-performing models.
\end{finding}

While the overall scores appear promising, with most models achieving over 60\%, a closer examination of completely correct answers (Figure~\ref{fig:binary}) reveals a less optimistic picture. We define an answer as completely correct if the model receives full credit in at least one of three trials. Under this criterion, all models struggled significantly, with the best performer (Grok 4) answering only 3 of 15 questions with full credit. When we apply a stricter definition, requiring full credit across all three trials, the results are even weaker: Grok 4 and Claude Opus 4 each answered 2 questions correctly, while all other models answered at most 1 question completely correctly, with most achieving 0.

\begin{finding}
Models rarely responded with completely correct answers. Under stricter criteria (full credit across all trials), the best models answered at most 2 questions correctly, with most models achieving 0 or 1.
\end{finding}

\noindent\textbf{Performance Across Programming Languages.}
As shown in Figure~\ref{fig:overall}, models performed best on Java (66.86\%) and C++ (63.21\%), but struggled with Python (50.44\%). We also report each languages deviation from the average in Figure~\ref{fig:lang}. Nearly all (19 of 20) models scored below their average on Python, with only Claude Opus 4 showing positive deviation. Some models maintained balance across languages (Claude Opus 4.1), while others (GPT-4.1) show extreme strengths and weaknesses. More specialized models tend to perform worse overall.  However, top performers span a range of language deviation variance, suggesting language balance is not the sole determinant of quality.

\begin{finding}
Models struggled on Python questions (avg: 50.44\%), with nearly all underperforming their average. Although there is a correlation between language deviation and performance, top performers vary widely in cross-language consistency, from balanced (Grok 4, variance 23.3) to large performance gaps (GPT-5, variance 101.5).
\end{finding}

\noindent\textbf{Performance Across Question Type.} 
In Figure~\ref{fig:questiontype}, we present a heatmap of performance on each question, grouped by the question types described in Section~\ref{sec:benchmark} and sorted by score. We find that, on average, models perform best on Library Behavior questions (63.7\%), and struggle on Project Behavior questions (55.0\%). In other words, LLMs are better at answering contextualized questions about API usages than in project code. We find that OpenAI models (68.8\%) significantly outperform other models (53.8\%) on Performance questions, with all five models scoring above 65\% and gpt-oss-20 leading with 73.4\%. Anthropic models (71.2\%) show an advantage on Library Behavior questions compared to all other models (61.8\%), with Claude Opus 4.1 and Claude Opus 4 achieving 82.1\% and  81.5\% respectively.
\begin{finding}
Models perform best on Library Behavior questions (63.7\%), and struggle most with Project Behavior questions (55.0\%).
\end{finding}

%

\subsection{RQ2: Resource Usage vs Performance }
In this section, we provide a discussion of the resource consumption of each model we evaluate and the relation to its overall performance. For proprietary models, we analyze the cost in USD. For open source modules, we analyze their size in number of parameters. We recognize that some of the non-proprietary models are not \textit{truly} open source in their training data and processes, but for this analysis we consider any model that we can run locally to be open source.

\begin{figure*}[t]
\vspace{-10pt}  
\centering
\begin{subfigure}[b]{0.47\textwidth}
    \includegraphics[width=\textwidth]{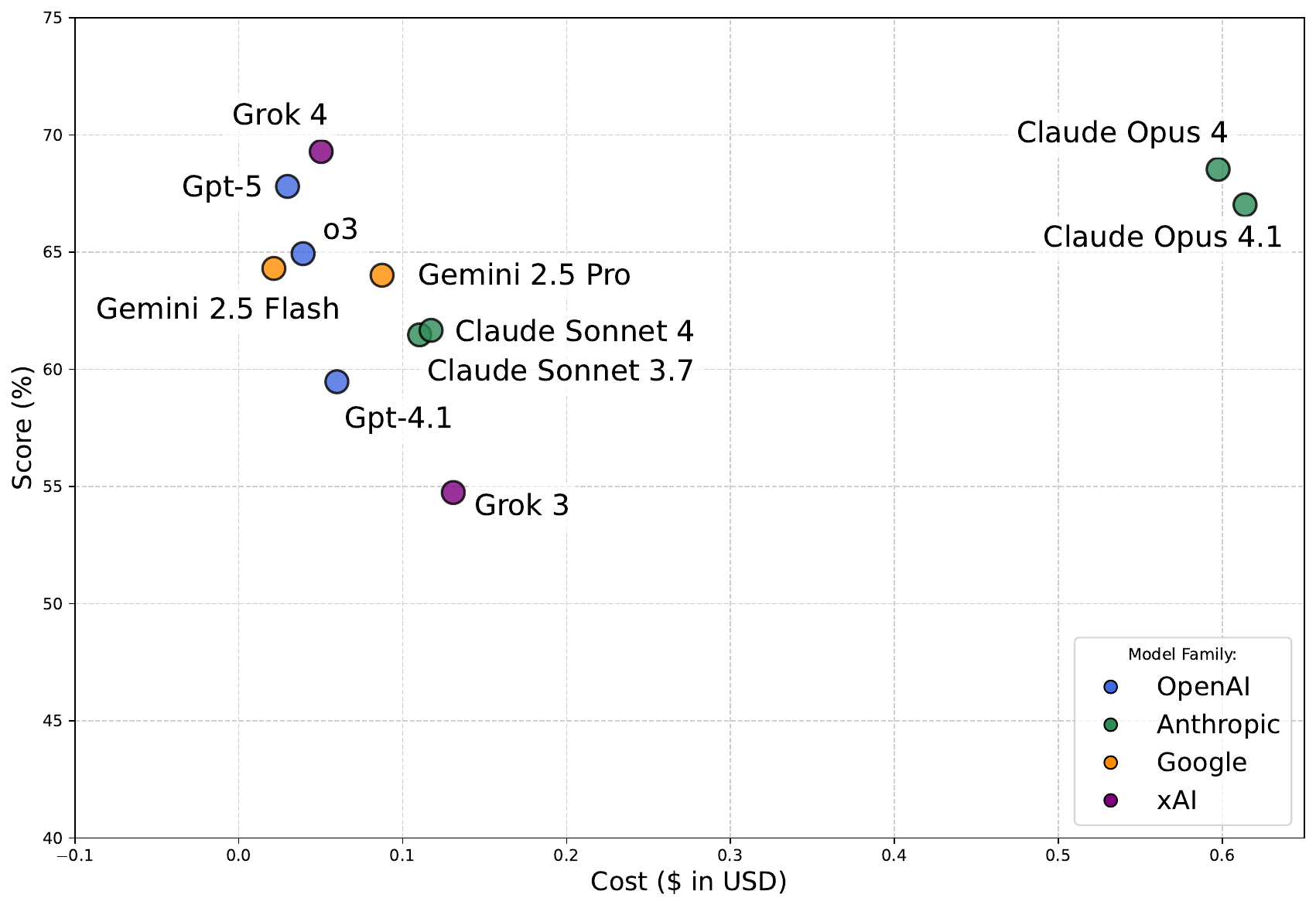}
    \caption{Performance vs Cost.}
    \label{fig:cost}
\end{subfigure}
\hfill
\begin{subfigure}[b]{0.47\textwidth}
    \includegraphics[width=\textwidth]{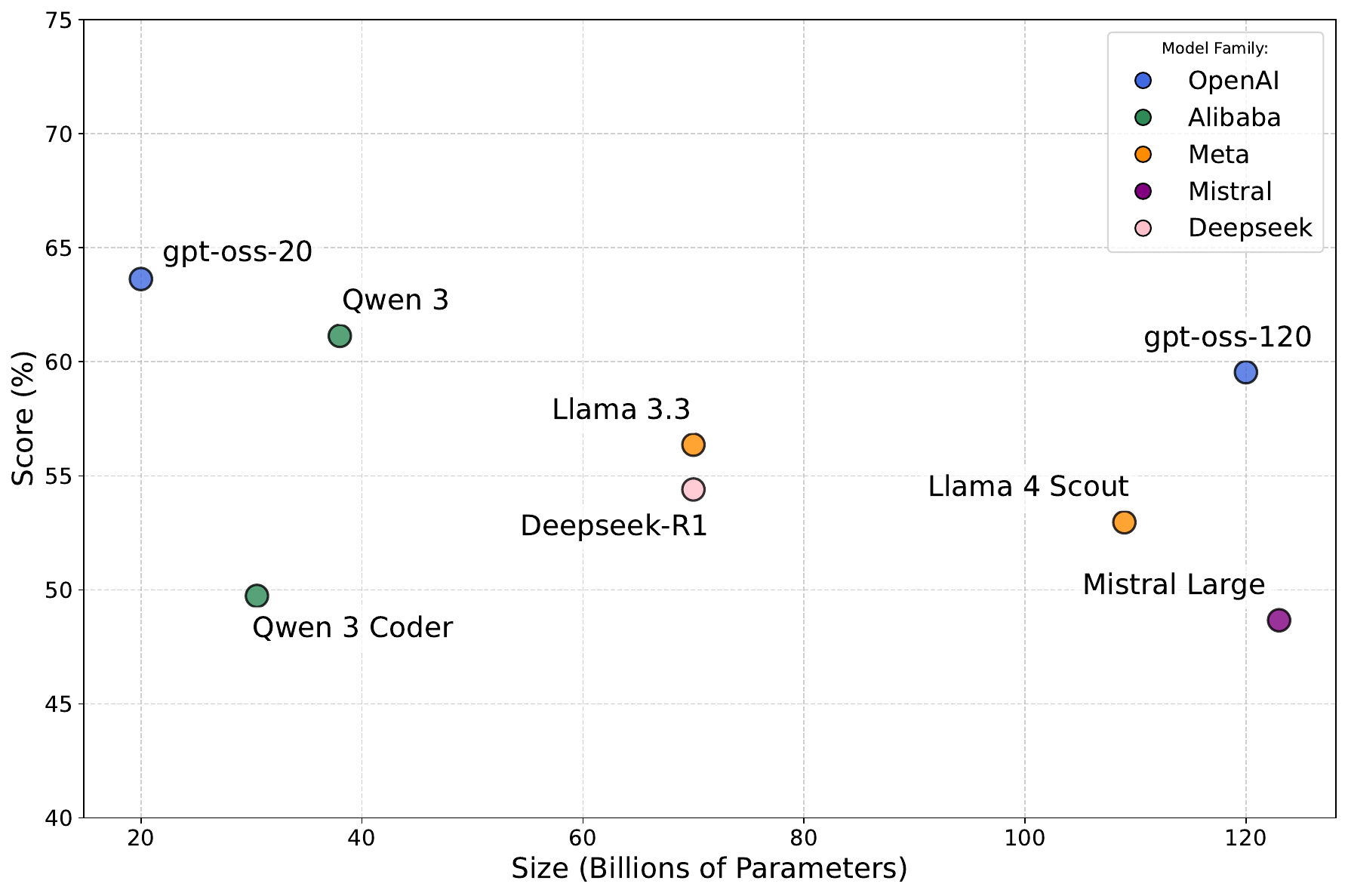}
    \caption{Performance vs Model Size.}
    \label{fig:size}
\end{subfigure}
\caption{RQ2: Resource Usage of Proprietary and Open Source Models.}
\Description{Performance vs Cost scatter plot for proprietary models shows Claude Opus models as high cost outliers. Performance vs Size scatter plot for open source models shows no correlation in size of model and score.}
\vspace{-10pt}  
\end{figure*}

\noindent\textbf{Performance Across Cost.} 
We calculate API costs based on input and output tokens averaged over 3 trials and plot cost versus performance in Figure~\ref{fig:cost}. Claude Opus models were high-cost outliers, with costs 1 standard deviation above the mean. The second-best performing model, Claude Opus 4, cost \$0.597, while the best performer, Grok 4, cost only \$0.05. To quantify the cost-performance tradeoff, we compute a ratio of cost to score (and multiply by 100) for each model. Gemini 2.5 Flash (0.033), GPT-5 (0.044), and o3 (0.060) achieved the lowest ratios, while Claude Opus 4.1 (0.916), Claude Opus 4 (0.872), and Grok 3 (0.239) had the highest. Thus, Claude Opus models deliver high performance at very high cost, whereas Grok 3 offers low performance (54.74\%) at low cost (\$0.13). Lastly, our evaluation included ``budget'' models Gemini Flash and Claude Sonnet, advertised as low-cost alternatives to flagship models Gemini Pro and Claude Opus. Gemini 2.5 Flash cost \$0.07 less than Gemini Pro while performing better (64.30\% vs. 64.01\%). Claude Sonnet models cost \$0.49 less than Claude Opus with only a 6.21 percentage point performance decrease.

\begin{finding}
Claude Opus models are significantly more expensive with minimal performance gains. Grok 4 and GPT-5 achieve comparable or better performance at 12x and 20x lower cost respectively.
\end{finding}

\noindent\textbf{Performance Across Model Sizes.} 
In Figure~\ref{fig:size} we present data on the size of open source models and their average score on RubberDuckBench.  We find that larger models do not correlate with higher scores. In fact, the highest performing open source model, gpt-oss-20B (63.63\%) was also the smallest (20B parameters) we evaluated, performing better than gpt-oss-120B (59.54\%).

\begin{finding}
Larger models do not correlate with higher scores. The highest performing open source model, gpt-oss-20 (63.63\%) was also the smallest model we evaluated.
\end{finding}
\subsection{RQ3: Hallucination in Responses}
\begin{figure}
\includegraphics[width=0.49\textwidth]{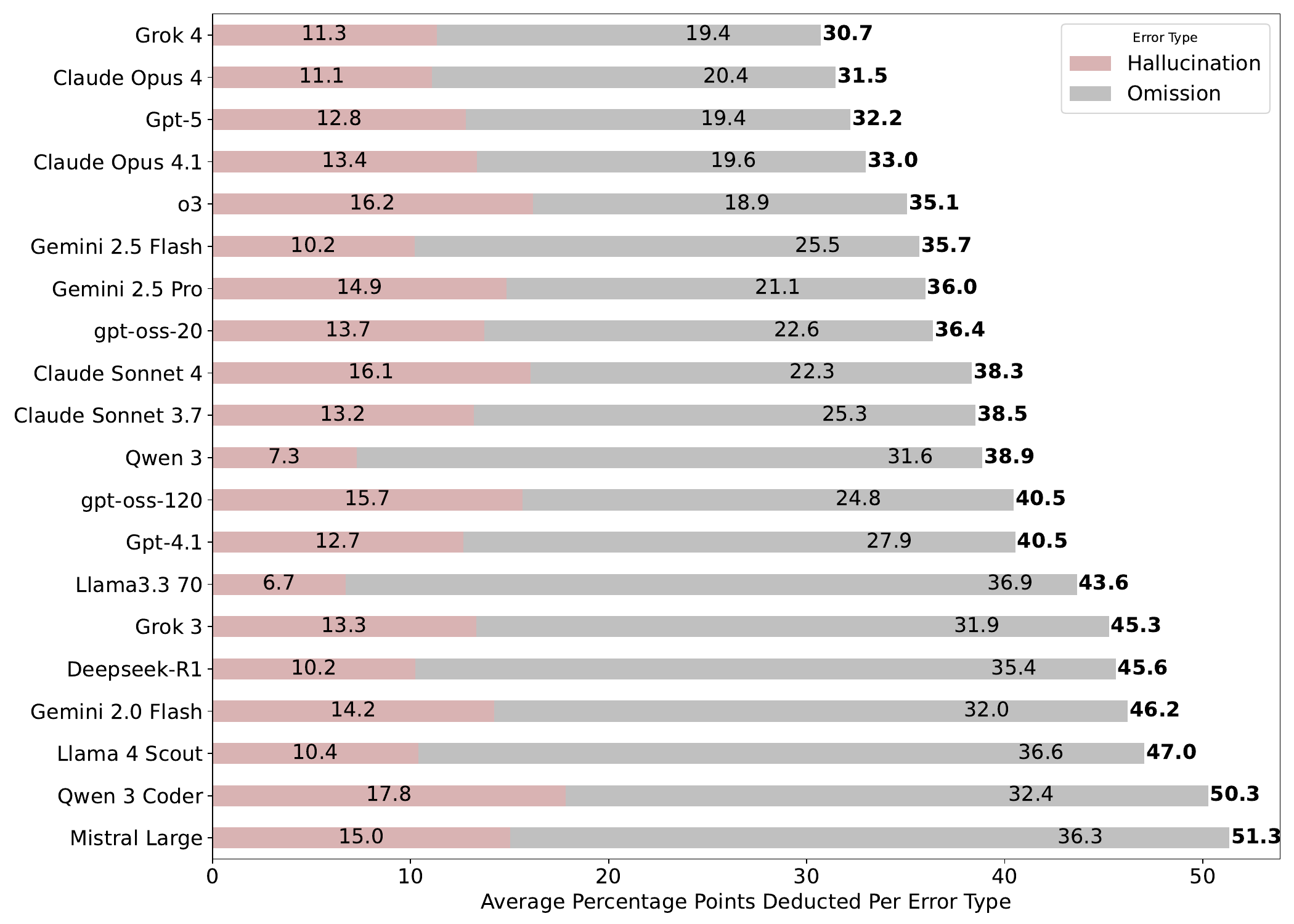}
\caption{Point Deduction Per Error Type.}
\label{fig:lies}
\Description{Breakdown of points deducted by hallucination or omission.}
\end{figure}

In this section, we study the frequency of hallucination in LLM responses. As discussed in Section~\ref{sec:curation:rubric}, our rubrics separately penalize hallucinating project or library information more heavily than omitting information. Figure~\ref{fig:lies} shows the percentage points deducted for hallucinations versus omissions. We find that, Qwen 3 Coder (17.8\%), Claude Sonnet 4 (16.1\%), and gpt-oss-120B (15.7\%) receive the most percentage points deducted for hallucinations. We also measured the total number of answers with hallucinations per model. Although o3 was one of the best performing models, it lied in 10 of its 15 answers (67\%). On average, the LLMs we studied provided answers with hallucinations in over half the questions (58.3\%). The models that hallucinated the least were Grok 4 (6 questions, 40\%), DeepSeek-R1, Qwen3, and Llama3.3 70 (7 questions, 46\%). Except for top-performing Grok 4, these models achieved lower overall scores, with most deductions due to omitting information rather than hallucinating.

\begin{finding}
LLMs gave responses with hallucinations to 58.3\% of questions on average. Even high-performing models like o3 hallucinated frequently (in responses to 10 questions, 67\%).
\end{finding}

\section{Conclusion}
In this paper, we present RubberDuckBench: a benchmark of 15 contextualized questions for AI coding assistants. Grok 4 is the highest performing (69.29\%) but does not exhibit pairwise significant superiority over the next 12 best performing models. We find that models rarely responded with completely correct answers, with the best models only answering at most 2 questions completely correctly across all trials. Furthermore, models often hallucinate with lies in 58.3\% of responses on average. Cost analysis reveals no correlation between expense (API pricing or parameter count) and performance. We recognize that the following threats to validity exist. Firstly, PR comments were filtered and rephrased by an LLM which may induce a bias. We attempt to mitigate this by manually inspecting each flagged comment and rephrasing when necessary. Secondly, as the projects are popular open source repositories, they may be members in a given LLMs training set. Although the code may have been seen prior, the questions in the current format, have not been trained on by any of the evaluated LLMs. We intend this benchmark to be a target for future research in trustworthy and correct AI coding assistants. 

\bibliographystyle{ACM-Reference-Format}
\bibliography{refs}



\end{document}